# A Novel Reinforcement Learning Routing Algorithm for Congestion Control in Complex Networks

S. yajadda , F. Safaei

Faculty of Computer Science and Engineering, Shahid Beheshti University, Tehran, Iran

s.yajadda@mail.sbu.ac.ir, f_safaei@sbu.ac.ir

## Abstract

Despite technological advancements, the significance of interdisciplinary subjects like complex networks has grown. Exploring communication within these networks is crucial, with traffic becoming a key concern due to the expanding population and increased need for connections. Congestion tends to originate in specific network areas but quickly proliferates throughout. Consequently, understanding the transition from a flow-free state to a congested state is vital. Numerous studies have delved into comprehending the emergence and control of congestion in complex networks, falling into three general categories: soft strategies, hard strategies, and resource allocation strategies. This article introduces a routing algorithm leveraging reinforcement learning to address two primary objectives: congestion control and optimizing path length based on the shortest path algorithm, ultimately enhancing network throughput compared to previous methods. Notably, the proposed method proves effective not only in Barabási-Albert scale-free networks but also in other network models such as Watts-Strogatz (small-world) and Erdös-Rényi (random network). Simulation experiment results demonstrate that, across various traffic scenarios and network topologies, the proposed method can enhance efficiency criteria by up to 30% while reducing maximum node congestion by five times.

## 1. Introduction

Network theory stands out as a pivotal interdisciplinary field, offering solutions to pressing issues in public health, technology, the environment, socioeconomics, and more, that confront humanity today. Graphs are instrumental in modeling many complex real-world systems, including the Internet, metabolic networks, power grids, food webs, and transportation networks. The burgeoning field of network science aims to describe, predict, and control such intricate systems.

Efficiently studying, probing, and assessing complex networks and systems constitute a prestigious realm of interdisciplinary research, enlisting contributions from physics, mathematics, biology, social sciences, informatics, and various practical and theoretical disciplines. Among the primary challenges within this domain is the phenomenon of traffic and congestion. The onset of congestion not only diminishes network efficiency but can also lead to network loss. Hence, monitoring the network's state and understanding the transition from a free state to a congested state become imperative tasks in network analysis.

### 1.1 The Problem Statement

Congestion and traffic are inherent challenges in networks, capable of manifesting in any segment of the network and posing threats such as disrupted functions or complete collapse. Several factors contribute to network congestion. Network structure, despite having similar average degrees, nodes, and links, may exhibit different capacities. The routing of packets from source to destination is another critical factor influencing congestion. Efficient resource allocation, particularly in utilizing links and buffers, constitutes a third significant factor impacting congestion.

Congestion in a network has the potential to rapidly propagate, starting from a single point and swiftly enveloping the entire network, leading to loss and collapse. Consequently, prompt detection and prevention of congestion are imperative, with the effective functioning of the network hinging on avoiding congestion. Identifying potential congestion points becomes crucial for preemptive interventions and actions to forestall the emergence of congestion.

The objective of this article is to introduce a congestion control routing algorithm that not only preserves network integrity but also enhances its performance. The proposed method is based on the shortest path algorithm; however, it adjusts routes slightly during congestion control to ensure minimal disruption to network performance. Specifically, the aim is to offer a routing algorithm that not only ensures the proper functioning of the network but also optimally distributes the network load, minimizing the distance packets travel within the network.

The innovation lies in presenting a routing algorithm rooted in the shortest path algorithm. It dynamically updates critical paths using reinforcement learning and Q-learning at each step to prevent congestion. The algorithm adheres to the shortest path in low-traffic conditions and switches to alternative routes as network traffic intensifies. The primary focus is on soft strategies, providing a routing algorithm for optimal packet routing from source to destination. The core structure of the network and resource allocation remain unchanged.

The incorporation of reinforcement learning ensures accurate route updates, with one route from each node being updated in each step, balancing computational overhead. The manuscript is structured into six sections. Section 2 elucidates fundamental concepts, facilitating a better understanding of the article. Section 3 provides an overview of prior work, presenting a general classification of solutions in the field. Section 4 details the proposed method, while Section 5 evaluates and analyzes the results from experimental simulations. Finally, Section 6 summarizes the article's contents, concludes the findings, and outlines avenues for future research.

## 2. The Literature and Basic Concepts

The effective betweenness centrality involves calculating the betweenness centrality at each time step, allowing for the measurement of the ratio of packets passing through a specific node to the total number of packets in the network. This approach assesses the proportion of utilized packets from network nodes and the distribution of network paths [1, 2].

### 2.1 Machine Learning

Machine learning pursues the development of computational methods for "learning" from accumulated experiences. Traditionally, there are two fundamental types of learning in machine learning. The first is unsupervised learning, which focuses on unveiling inherent structures within data relationships. In this case, the learning process relies solely on the presented data, as no prior knowledge about the data is supplied. Key tasks of unsupervised learning include clustering, outlier detection, and pattern recognition.

The second type is supervised learning, aiming to infer data-related concepts using provided labeled samples, typically represented as the training set. The primary distinction between supervised and unsupervised learning lies in the fact that in supervised learning, the learner examines external information from the training set during the training phase to deduce the classifier hypothesis. In a classification task, this external information is incorporated into the learning process through classes or labels. The goal of classification is to create a prediction function that can generalize from the training set when applied to unseen data (the test set).

In contrast, the unsupervised learning model seeks patterns or trends in data and attempts to categorize them, grouping similar data while segregating data of different types into distinct clusters [3].

### 2.2 Q-Learning

Q-learning is a reinforcement learning technique that adheres to a particular policy for executing various actions in different situations through the learning of an action-value function. A notable strength of this method is its capacity to learn the mentioned function without requiring a specific environment model. Q-Learning endeavors to select optimal actions based on the present circumstances. This algorithm is categorized as off-policy. As the Q-learning function executes actions outside the current learning policy, the approach typically aims to maximize the total reward. The relationship defining the total reward is articulated as follows [4]:

$$Q(S_t, A_t) \longleftarrow Q(S_t, A_t) + \alpha[R_{t+1} + \gamma \max Q(S_{t+1}, a) - Q(S_t, A_t)] \qquad (1)$$

where $Q(S_t, A_t)$ on the left side of the relation represents the updated value of Q, and $Q(S_t, A_t)$ on the right side of the relation signifies the previous value of Q. The parameters in the equation include $\alpha$, the learning rate; $\gamma$, the discount factor; and $R_{t+1}$, the reward value.

## 3. Related Work

In this section, various methods and approaches to prevent congestion are introduced. The classification of these methods, the evolution of related theories, and their respective strengths and weaknesses are explored to provide a comprehensive understanding of congestion and its prevention strategies. Two primary approaches to control congestion, namely soft and hard strategies, will be discussed. Additionally, a third category known as resource allocation will be introduced. The subsequent discussion will offer a detailed description of each method, highlighting the specific aspects they address and emphasize.

### 3.1 Soft Strategies

In the realm of soft strategies, congestion solutions are explored, with a focus on determining the routing of network packets. These methods aim to be cost-effective [5, 6] by leveraging the properties of network nodes, the type of network topology, etc. The simplest routing approach in this category is the shortest path algorithm, where packets select the shortest distance between the source and destination as their route. However, a notable drawback of this method is that, in real networks and under high traffic, nodes with high degrees (hubs) quickly become congested [7, 8]. Another proposed method, presented by Yan et al. [9], introduces the concept of an "efficient path." In this approach, the sum of the relation defined below should be minimized for each path between two nodes $i$ and $j$, with the value of the $β$ parameter.

$$L(P(i \rightarrow j) : \beta) = \sum_{i=0}^{n-1} k(x_i)^\beta \qquad (2)$$

Furthermore, when the exact value is determined for two different paths between nodes $i$ and $j$, one is randomly selected [9]. A similar scheme has been proposed, where, despite employing the same Equation (2), packets are prioritized for transmission [10].

Another method, known as the "aware traffic strategy," has been introduced [11, 12]. In this approach, a maximum time $<T_{max}>$ is assigned for each packet in the network, representing the duration it takes for the packet to reach the destination from the source. According to this method, the traffic capacity is increased compared to the shortest path method, aiming to mitigate congestion gradually as traffic in the network intensifies.

Zhao et al. [13] proposed a method based on two network models: random networks, regular networks, symmetric trees, and scale-free networks. A comparison of the results between the two models revealed that in the first model, where the capacity of each node is equal to its degree, scale-free and random networks perform better in preventing congestion compared to other networks. In the second model, where the processing capacity of a node is higher with increased centrality, there is a relative improvement in the performance of random and scale-free networks, and the performance of regular and tree networks is also significantly enhanced [13].

In [14], a solution based on local information is presented. Packets are routed in the network based on the information of neighboring nodes, utilizing Equation (3). If the next node of the packet corresponds to the destination type, it is directed straight to that node; otherwise, the direction is determined by calculating the probability through Equation (3), favoring nodes with higher probabilities.

$$\Pi_i = \frac{k_i^\alpha}{\sum_j k_j^\alpha} \qquad (3)$$

In the above equation, k represents the node degree, and α is a variable parameter. In this method, the buffer length of the nodes is considered unbounded, and the first-in-first-out policy is applied. A crucial rule of this method is that no packet can traverse the same path more than twice.

Holm [15] investigated how the speed of traffic flow in the network is influenced by the network structure, introducing three routing methods. The first method, abbreviated as RW (random walk), involves packets randomly traversing nodes based on the shortest path until reaching the target node. This method performs well at low traffic and exhibits acceptable performance when the network load is negligible. The second method is obstacle avoidance, abbreviated as DO. In this approach, the packet determines the path and target node through random walking. When encountering a node with other packets waiting for routing, the packet bypasses that node and selects an alternative path. This method functions similarly to the random walk method under low-traffic conditions. The third method is the wait-on-obstacle approach. Here, the packet utilizes the random walk method and, when facing a congested node, waits until the traffic at that node subsides before continuing its route. The drawback of this scheme is the potential for long waiting times, particularly in scale-free networks where some nodes have high degrees, potentially leading to network deadlock.

De Martino et al. [16] proposed a method as a compromise between structure-based and traffic-based strategies. They argued that traffic control in homogeneous networks is futile and adds little value, but it can enhance the network's performance in heterogeneous networks. This perspective is based on the assumption that homogeneous networks have similar degrees of distribution and traffic, which is not reflective of real-world networks. Thus, investigating homogeneous networks may offer limited practical advantages. Moreover, the authors believed that congestion control should be dynamic and tailored to the network load.

Yang et al. [17] introduced a strategy based on the shortest path and betweenness centrality. In this method, the weight of each edge is determined using the following relation. Subsequently, the least weight path between the source and destination is selected for packet navigation.

$$W_{ij} = (1 + B_i)^\alpha + (1 + B_j)^\alpha \tag{4}$$

In the above equation, the parameter $W_{ij}$ represents the weight between two nodes $i$ and $j$. The variable quantities α and $B_i$ indicate the centrality of betweenness for node $i$. Another method reported in [17] simultaneously considers two parameters: degree centrality and betweenness centrality.

$$W_{ij} = [(1 + B_i) \cdot k_i]^\alpha + [(1 + B_j) \cdot k_j]^\alpha \tag{5}$$

In the above relation, $k_i$ depicts the degree of the node. This method's advantage is increasing the network capacity and reducing the packet transmission time. Additionally, researchers [17] have claimed that their proposed method is scalable and will exhibit an optimal response on different networks.

Echagüe et al. [18] proposed a method with two primary and essential goals: preventing the network from collapsing due to congestion and increasing the tolerance of the network. They aim to ensure effective communication between nodes while utilizing the shortest possible path for intra-network communication. In their proposed method, the weight of each node and one of its neighboring nodes is calculated using the following equation, and more weight will be assigned to the link's weight. Based on these weights, the shortest path for communication is considered.

$$W_i^t = (1 + C_i^t)^\gamma \tag{6}$$

where γ is a variable quantity, and $C_i^t$ represents the amount of congestion in node $i$ at time $t$, which can be calculated using the following equation:

$$C_i^t = \frac{\rho B_i^t}{n - 1} \tag{7}$$

in which, ρ represents the packet generation rate in the network, $B_i^t$ is the betweenness centrality of node $i$ at time $t$, and $n$ represents the number of nodes in the network. The advantages of this method include the dynamics of the packet transmission path and attention to congestion in the nodes, making it an efficient approach to prevent congestion in the network [18].

A recent report [19] by the same authors investigated the impact of two crucial factors on congestion prevention strategies, highlighting the effect of each on network performance. These factors involve updating all routes simultaneously and using the information of a group of nodes to calculate suitable routes.

Simultaneously updating all routes may increase the probability of network congestion due to updating multiple routes from the same node. Therefore, a logical solution to this problem could be to update only one route from each node simultaneously. Additionally, the influence of the betweenness centrality criterion should be considered when utilizing the information of nodes. Nodes with high betweenness centrality, referred to as "known nodes" in this research [19], have a significant effect on congestion occurrence. By distributing traffic among nodes with much less betweenness centrality, known as "unknown nodes," it is possible to prevent congestion [19].

## 4. The Proposed Method

In this section, we will first examine the challenges of previous methods and strategies. Subsequently, we will delve into how the proposed method, the novel routing algorithm, operates, demonstrating its effectiveness in reducing congestion in the network.

When reviewing the soft strategies discussed in Section 3.1, certain areas for improvement become evident. One notable challenge is their topology-based nature, where existing methods are tailored to specific networks with predefined characteristics, limiting their applicability to practical real-world networks.

Another critical consideration is the computational overhead associated with these methods. Introducing algorithms with high calculations and complexity poses challenges, particularly in handling large networks. While these methods may contribute to a relative reduction in network congestion, they may also introduce new computational challenges.

The proposed method aims to address these issues and control network congestion effectively. Leveraging reinforcement learning, widely employed across various scientific fields, proves to be a significant asset in this context.

The roles of betweenness centrality and degree centrality are pivotal in understanding network congestion. Nodes with high betweenness centrality are susceptible points for congestion in the network. Additionally, nodes with high degree centrality foster increased connections, contributing to congestion. Specifically, the two factors of degree centrality and betweenness centrality are considered to enhance the understanding of congestion in nodes. By leveraging these factors, we can guide packets effectively through the network, suggesting optimal routes from source to destination (see Figure 1).

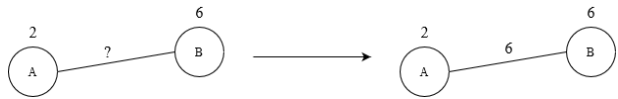

**Figure 1:** Determining the weight of links in the proposed method.

For each packet's routing, the critical goals of this method are to utilize the minimum possible path to reach the destination, enhance network throughput, and increase network capacity. The proposed method aims to use the shortest route in low traffic conditions, and as network traffic increases, the network load is distributed away from nodes prone to congestion.

In this method, each node serves three primary roles in the network: it can act as the source or destination of a packet, function as an intermediate router, and direct the packet to the next node. Each node has the capacity to process packets, assumed to be equal to one, implying that each node can handle one packet per unit of time. The packet generation rate in the network, represented by ρ, denotes the probability of packet generation in each network node, ranging between zero and one.

The routing method advocates for the least weighted route between the source and destination by considering the weight of the link between nodes. According to this method, the weight is calculated for each node using Equation (8). Subsequently, as routing is based on the link weight, a higher value is assigned to the link weight between two nodes, A and B.

$$W_i^t = (1 + C_i^t)^\gamma \qquad (8)$$

where $\gamma$ is a tunable parameter, and $C_i^t$ represents node congestion, which can be determined using the following equation:

$$C_i^t = \frac{\rho k_i B_i^t}{n-1} \tag{9}$$

In which $k_i$ is the degree of node $i$, $B_i^t$ represents the betweenness centrality of node $i$ at time $t$, and $n$ is the number of nodes in the network. The reason for using effective betweenness centrality is to examine the dynamic behavior and performance of the network and make appropriate decisions accordingly.

Updating the routes is also one of the essential features of this method, providing a dynamic view of traffic flow and network behavior for making informed decisions. However, updating all routes poses two problems. Firstly, it leads to high computational overhead, as all link weights in the network must be updated continuously, requiring a significant number of calculations. Secondly, simultaneous updating of all routes can cause congestion to shift from one point to another in the network.

To address these issues, a compromise is proposed by updating only one route from each node at each time step, reducing the computational load significantly. The challenge lies in how to select this path. While previous research chose the route randomly, the proposed method uses reinforcement learning for path selection. The weight of each node and the chosen node are calculated using Q-learning and the Q-table based on Equation (9). The node with a higher value is selected as the weight of the link between two nodes.

In reinforcement learning, the agent makes decisions by monitoring network conditions and considering the learning rate. Using Q-learning, only one path from each node is updated at each time step. The agent decides, based on the learning rate, network conditions, and traffic, which path to update. The agent's reward in this method is calculated from the reciprocal weight of each link between two nodes, as obtained from Equation (8). Q-learning, considered an off-policy method, aims to maximize the reward, deciding which path is more suitable for updating based on network conditions, without the need for external intervention as in other methods like SARSA [2].

## 4.1 Implementing the Proposed Method

In this study, the Python language and related libraries were employed to simulate the proposed algorithm. Various libraries, such as NetworkX, were used for generating network topology and node information. Additionally, Numpy and OpenAI libraries were utilized to implement reinforcement learning. The simulation of the proposed algorithm involved three scenarios:

1. Scale-free network Barabási-Albert model: This scenario employed a Barabási-Albert model with 256 nodes and *m*=3 (the number of links of the newly arrived node connected to existing nodes). The simulation began with a zero packet generation rate, increasing by 0.001 every 100 time steps until reaching 0.99, resulting in 10,000 time units. Packets' source and destination were randomly selected, and any packet reaching its destination was removed from the network. Node buffer capacity was assumed to be one, with a learning rate of 0.5 and a discount factor of 0.8.
2. Watts-Strogatz and Erdös-Rényi small-world models: In the second scenario, the efficiency of the proposed method was evaluated using the Watts-Strogatz and Erdös-Rényi small-world models. For the Watts-Strogatz model, a network with 64 nodes, a regular degree of 3, and a rewiring probability of 0.5 was simulated. For the Erdös-Rényi model, a graph with 64 nodes and a connectivity probability of 0.5 was assumed. Node buffer capacity was set to one, with a learning rate and a discount factor of 0.5 and 0.8, respectively.
3. Scale-free, small-world, and random models with Poisson process: In the third scenario, similar to the previous two scenarios, simulations were conducted for the scale-free, small-world, and random models. However, instead of randomly distributing packets, the Poisson process was used to generate and distribute packets.

Table 1 provides a summary of the model assumptions, network characterizations, and traffic patterns employed in the simulation experiments.

Table 1: Model assumptions and characteristics of the underlying networks

| Network model | Graph type | # nodes | # edges | Leaning rate | Traffic pattern |
|---|---|---|---|---|---|
| Barabasi-Albert | Undirected | 256 | 759 | 0.5 | Random |
| Erdös-Rényi | Undirected | 64 | 992 | 0.5 | Random |
| Watts-Strogatz | Undirected | 64 | 64 | 0.5 | Random |
| Barabasi-Albert | Undirected | 64 | 183 | 0.5 | Poisson |
| Erdös-Rényi | Undirected | 64 | 992 | 0.5 | Poisson |
| Watts-Strogatz | Undirected | 64 | 64 | 0.5 | Poisson |

In this section, we assess the effectiveness of the proposed method and evaluate its ability to achieve predefined objectives. The goals include efficiently controlling congestion, ensuring packets continue to use the shortest path, improving network throughput, increasing network capacity, and distributing the network load evenly across various points. The implementation of the proposed method utilized the Python language, leveraging its libraries and the powerful NetworkX library for certain aspects.

Throughout each simulation stage, we collected and analyzed the following statistics to measure and evaluate the effectiveness of the proposed method:

1. *Average Path Length*: This indicator represents the average path length taken by packets during each period. We investigated the impact of different methods on this index.

2. *Maximum Betweenness Centrality*: This parameter indicates the concentration and centrality of the load in specific points of the network, highlighting hub nodes. These nodes are more susceptible to congestion, and the proposed algorithm aimed to minimize this index.

3. *Throughput*: Throughput measures the percentage of packets reaching their destination within the network during a specific period and under a particular traffic pattern. Increased throughput indicates more successful packet delivery within the network.

4. *Maximum Node Congestion*: In each period, we monitored and analyzed the maximum congestion level calculated from Equation (9). This assessment helps track the emergence of congestion in network nodes and aims to prevent it as much as possible.

## 5. Results and Discussion

In this section, we analyze the simulation results of the first scenario, comparing our proposed method with the approach by Echagüe et al. [18] to assess the achievement of desired goals. The following comparisons are made:

- *Average Path Length* (Figure 2(a)): The average length of the path taken by packets in the network to reach the destination is plotted against the packet generation rate. The proposed method, based on Equation (8) and the proposed method based on reinforcement learning, is compared with the method reported in [18]. The results indicate that the proposed method, using reinforcement learning, outperforms the other two methods by about 10%, leading to a more efficient average path length criterion.

- *Maximum Betweenness Centrality* (Figure 2(b)): The network's maximum centrality of betweenness is examined concerning the packet generation rate. Simulation results reveal a significant improvement in the method based on reinforcement learning compared to the other two methods. This improvement is characterized by a balance in load distribution across the network, resulting in a 10% reduction in the maximum centrality of betweenness.

- *Maximum Node Congestion* (Figure 3(a)): The maximum node congestion in the network is measured and compared for the three methods. The results show that the proposed method has reduced the maximum node congestion by approximately 50%. The increase in the packet generation rate is associated with a more distributed and balanced load in the network, preventing congestion.

- *Network Throughput* (Figure 3(b)): The simulation results for the three methods are compared based on network throughput. The network throughput rate for the proposed method has improved by about 15% compared to the method reported in [18].

These comparisons collectively indicate the effectiveness of the proposed method in achieving the specified goals, demonstrating improvements in average path length, betweenness centrality, node congestion, and network throughput.

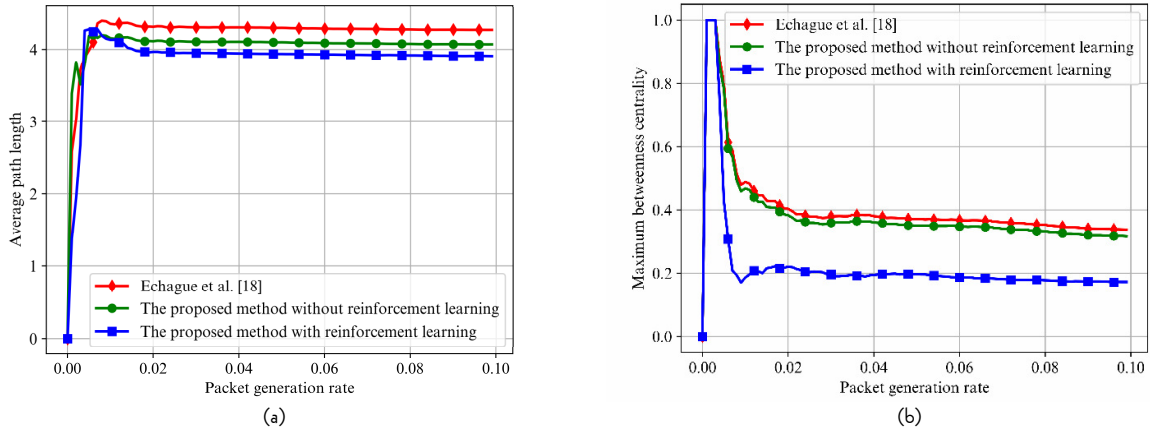

**Figure 2:** (a) Comparison of Average Path Length with Respect to Packet Generation Rate: This plot illustrates a comparison between three methods—Echagüe et al. [18], the proposed method without reinforcement learning, and the proposed method equipped with reinforcement learning—based on the average path length criterion. The results indicate that the third approach, incorporating reinforcement learning, outperforms the other two methods, achieving approximately a 10% improvement; (b) Comparison of Maximum Betweenness Centrality with Respect to Packet Generation Rate in the Barabasi-Albert Network Model: This chart presents a comparison of three methods—Echagüe et al. [18], the proposed method without reinforcement learning, and the proposed method with reinforcement learning—in terms of the maximum betweenness centrality criterion. The simulation results reveal a significant enhancement in load distribution and balance across the network with the adoption of reinforcement learning, leading to a 10% reduction in maximum betweenness centrality.

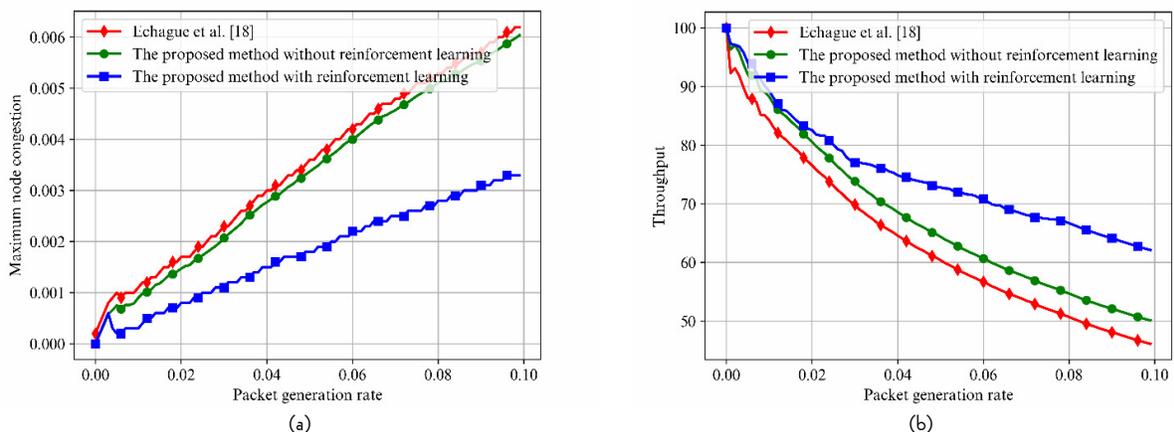

**Figure 3:** (a) Comparison of Maximum Network Congestion: This diagram depicts a comparison among three methods—Echagüe et al. [18], the proposed method without reinforcement learning, and the proposed method with reinforcement learning—in terms of maximum network congestion. The horizontal axis represents the packet generation rate, and the vertical axis indicates the maximum node congestion in the Barabasi-Albert network model; (b) Comparison of Throughput (Percentage): This chart illustrates a comparison of throughput percentages with respect to packet generation rate in the Barabasi-Albert network model for three methods—Echagüe et al. [18] under the random traffic model, the proposed method without reinforcement learning, and the proposed method based on reinforcement learning.

From the analysis of numerical results depicted in the graphs of this figure, it is evident that the proposed method, incorporating reinforcement learning, achieves a reduction in the average length of the route compared to method [18]. This improvement is attributed to the novel algorithm for weighting network links and the intelligent updates facilitated by reinforcement learning. Additionally, the proposed method enhances load distribution in the network, leading to a more balanced distribution without concentration in hub nodes. Consequently, the network throughput sees a notable increase of 15%. These findings collectively highlight the superior performance of the reinforcement learning-based method, especially when compared to method [18].

In the second scenario of the simulation, various complex network models were employed, and the effectiveness of the proposed method was measured and evaluated. Figure 4(a) illustrates the average path length index in the Erdös-Rényi random network model for three methods: method [18], the proposed method without reinforcement learning, and the proposed method based on reinforcement learning. The average path length in all three methods remains similar, hovering around 2.7 steps. Figure 4(b) displays the maximum betweenness centrality, showcasing a superior performance for the proposed method based on reinforcement learning with a reduction of approximately 30%. Furthermore, Figure 4(c) demonstrates a significant reduction in node congestion, dropping more than threefold from the maximum network congestion. However, Figure 4(d) reveals that the throughput index remains nearly identical for all three methods.

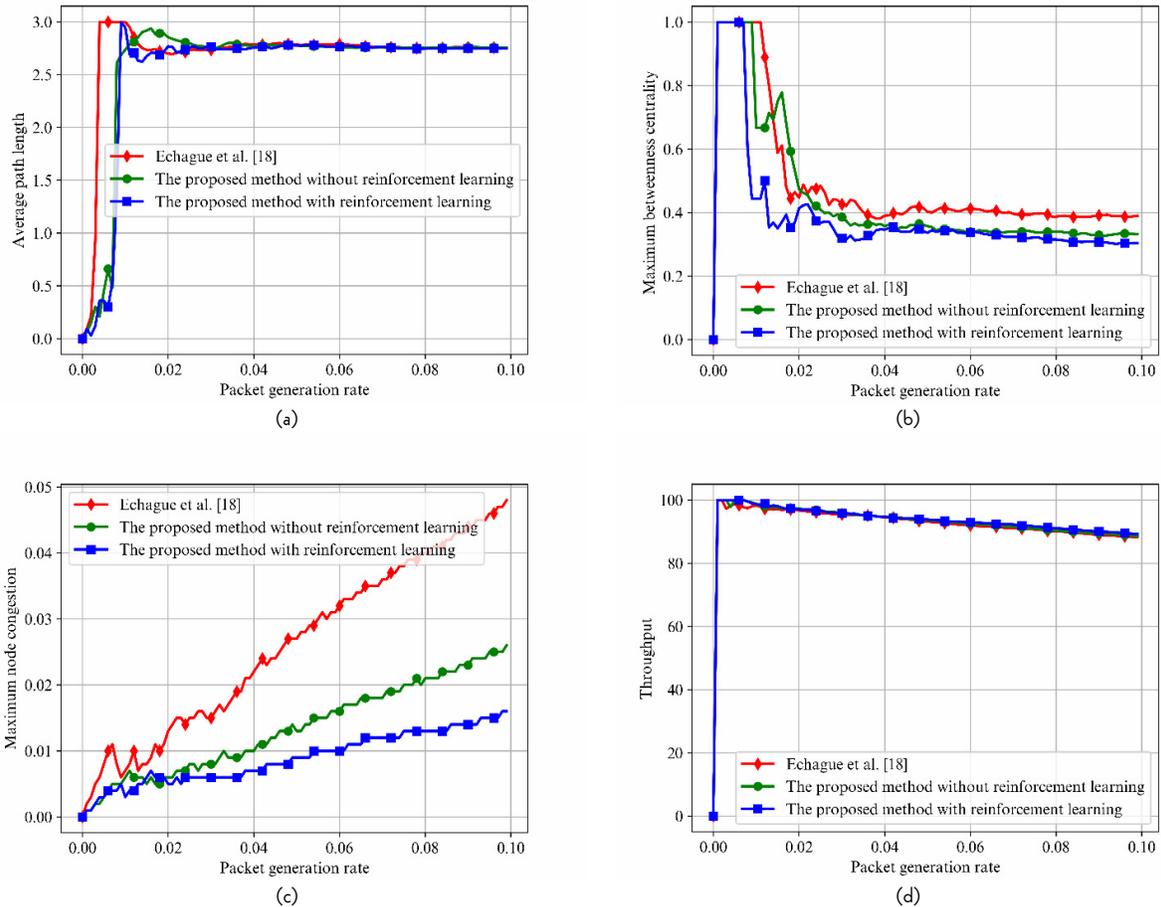

**Figure 4:** (a) Comparison of the average path length; (b) maximum betweenness centrality; (c) maximum node congestion, and (d) throughput (percentage) concerning the rate of packet generation in the Erdös-Rényi stochastic model network for Echagüe et al. [18] under the random traffic pattern, the proposed method without reinforcement learning, and the proposed method based on reinforcement learning. The number of nodes is assumed to be 64, and the probability of connectivity is assumed to be 0.5.

Figure 5(a) depicts the average path length for the Watts-Strogatz small-world model [20]. The number of nodes is 64, the initial graph is regular with a degree of 6, and the rewiring probability is assumed to be 0.5. It can be seen that the average path length has decreased by 20% in the proposed method, which is almost the same in both proposed methods. Moreover, in Figure 5(b), the maximum betweenness centrality in the proposed method based on reinforcement learning has decreased by 25%. However, focusing on Figure 5(c), we realize that due to the use of the proposed method based on reinforcement learning, the maximum congestion in the nodes is higher than in the other two methods. Still, considering the reduction of maximum betweenness centrality and according to Figure 5(d), which shows the percentage of throughput in different methods, it is apparent that the increase in throughput of the proposed method was less than 5%.

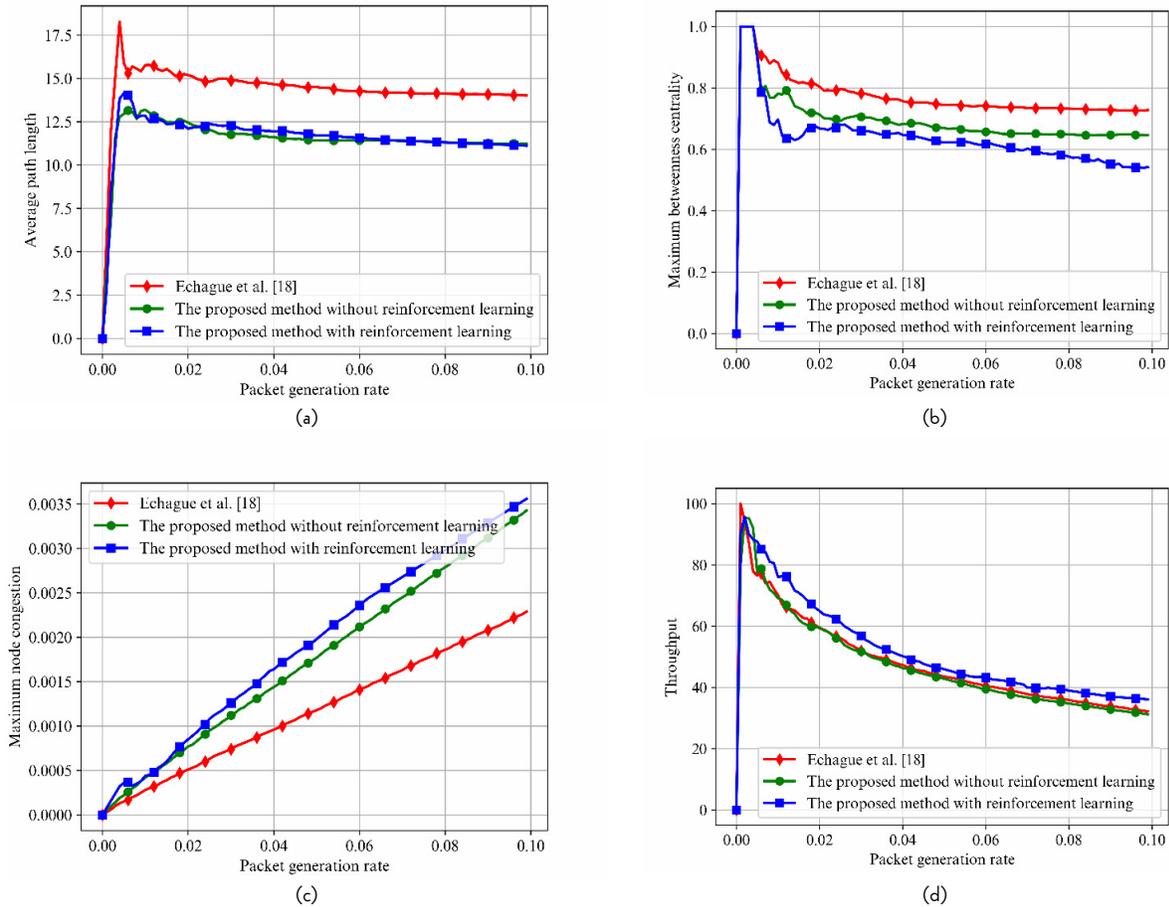

**Figure 5**: Comparison of average path length in the network with varying packet generation rates for the Watts-Strogatz small-world model across three methods under random traffic patterns. The network consists of 64 nodes, with an initial graph being regular with a degree of 6 and a rewiring probability of 0.5.

The empirical results analyzed thus far were predominantly based on the random traffic pattern in the network. However, the Poisson traffic pattern is widely utilized in modeling and evaluating network efficiency. Therefore, in the third simulation scenario, we employed the traffic pattern resulting from the Poisson process. Various numerical simulations were conducted on three network models, and the results obtained will be further investigated and analyzed.

Figure 6(a) illustrates the average path length for the Barabási-Albert scale-free network model [18, 19] under the Poisson traffic pattern for three different methods. The figure indicates that, with an increased rate of packet generation in the network and an enhanced learning process, the proposed method based on reinforcement learning has achieved an efficiency improvement of approximately 5% compared to other methods.

Figure 6(b) presents the maximum betweenness centrality index for this model under the aforementioned assumptions. While the methods of Echagüe et al. [18] and the proposed method without reinforcement learning exhibit similar functionality, the proposed method based on reinforcement learning has outperformed by about 15%. The results in Figure 6(c) also highlight the performance improvement of the proposed method. The proposed method without reinforcement learning has better managed the congestion issue in the network nodes compared to the other two schemes, particularly concerning the maximum node congestion in the network.

In the reinforcement learning-based proposed method, the maximum congestion in the nodes is five times less than in the research method [18]. Furthermore, the comparison of the throughput rate in the graphs of Figure 6(d) reveals that the throughput rate in the proposed method based on reinforcement learning is similar to the other two methods.

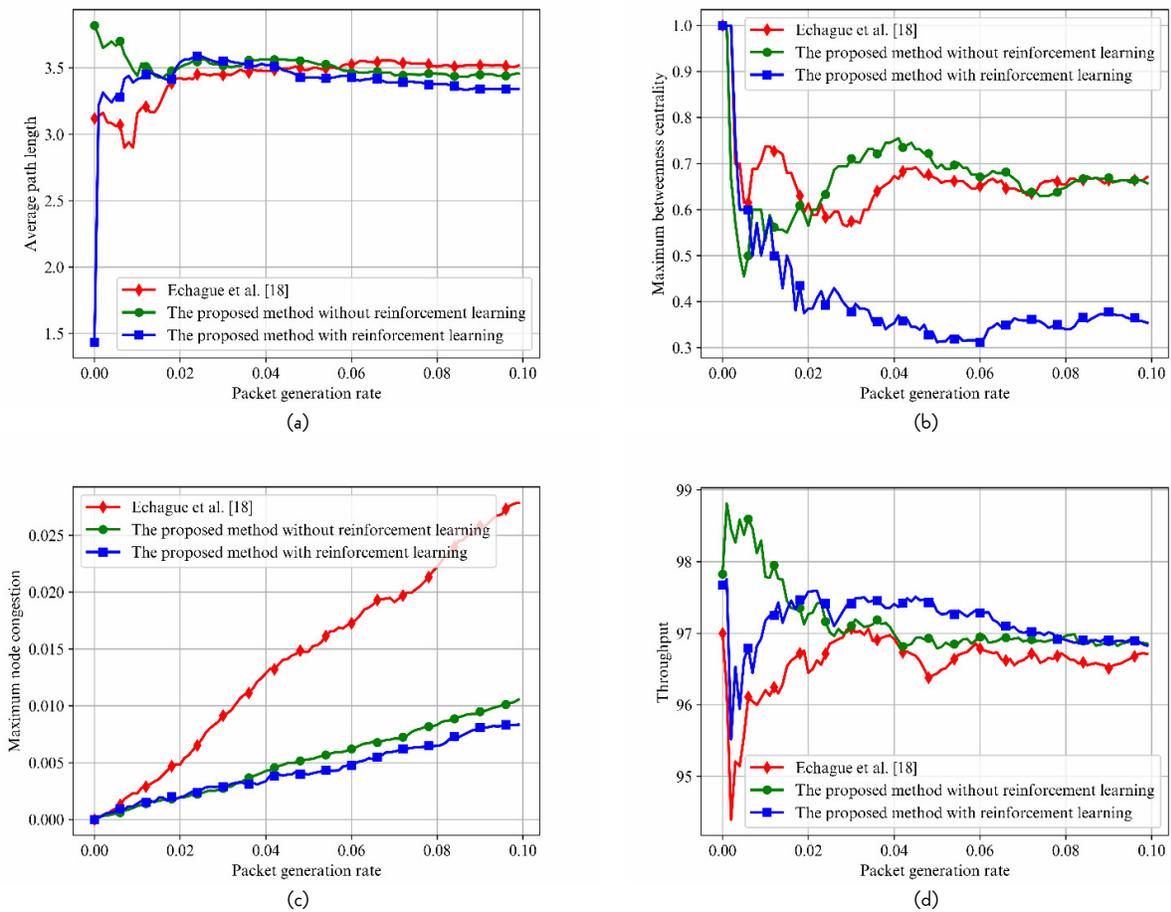

**Figure 6:** (a) Comparison of average path length; (b) maximum betweenness centrality; (c) maximum node congestion, and (d) throughput (percentage) concerning packet generation rate in the network using the Barabási-Albert model. The evaluation includes three different methods under the Poisson traffic pattern. The network comprises 64 nodes, with an initial core value of $m_0=3$, and each newly arrived node is assumed to have $m=3$ edges.

In the simulation experiments of the Erdös-Rényi random network model, illustrated in Figure 7(a), the average path length is comparable between the proposed method without reinforcement learning and the method presented in [18]. However, the proposed method based on reinforcement learning has achieved a 14% reduction in the average path length, as depicted in the figure. This optimization is also evident in Figure 7(b), where the maximum betweenness centrality in the reinforcement learning-based proposed method has decreased by approximately 50%, outperforming the approach reported in [18]. Figure 7(c) indicates that the maximum congestion of nodes is nearly the same for all three schemes. In terms of throughput, as shown in Figure 7(d), all three methods exhibit similar behavior.

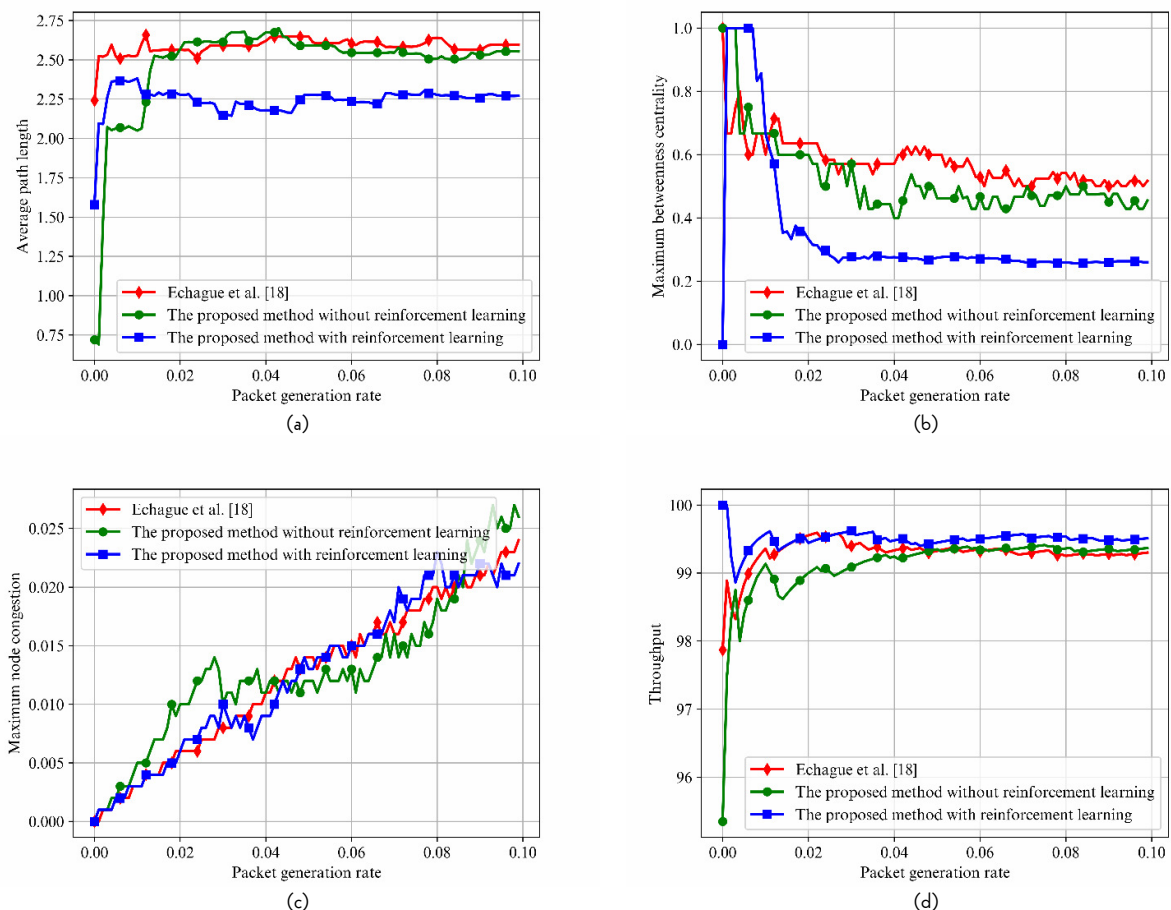

Figure 7: (a) Comparison of average path length; (b) maximum betweenness centrality; (c) maximum node congestion, and (d) throughput (percentage) concerning packet generation rate in the network using the Erdös-Rényi random network model. The evaluation involves three different methods under the Poisson traffic pattern, with 64 nodes and a connectivity probability of 0.5.

Furthermore, simulations were conducted under Poisson traffic conditions for the Watts-Strogatz small-world model. In Figure 8(a), the average path length is simulated for this model. The results indicate that the approach presented in [18] suggests longer paths for routing packets in the network, while the proposed method based on reinforcement learning selects optimal paths. In terms of network load distribution, the reinforcement learning-based proposed method outperforms the other two methods. Figure 8(b) illustrates the maximum betweenness centrality in all three schemes, showing that the proposed method with reinforcement learning performs more than 16% better than the method reported in [18]. The network throughput metric for all three methods is demonstrated in Figure 8(d), with an improvement of about 5%, indicating superior performance. Additionally, in the diagrams of Figure 8(c), the maximum congestion metric of the network nodes is included. The proposed method based on reinforcement learning is approximately five times faster than the study reported in [18] and four times faster than the proposed approach without reinforcement learning, resulting in less congestion in the network nodes. The numerical results obtained from simulations demonstrate that the proposed method, compared to the method proposed by Echagüe et al. [18], exhibits better performance and efficiency not only in the Barabási-Albert scale-free network model but also in other network models, namely, the Erdös-Rényi [21] and the Watts-Strogatz [20] small-world models, and under different network traffic patterns.

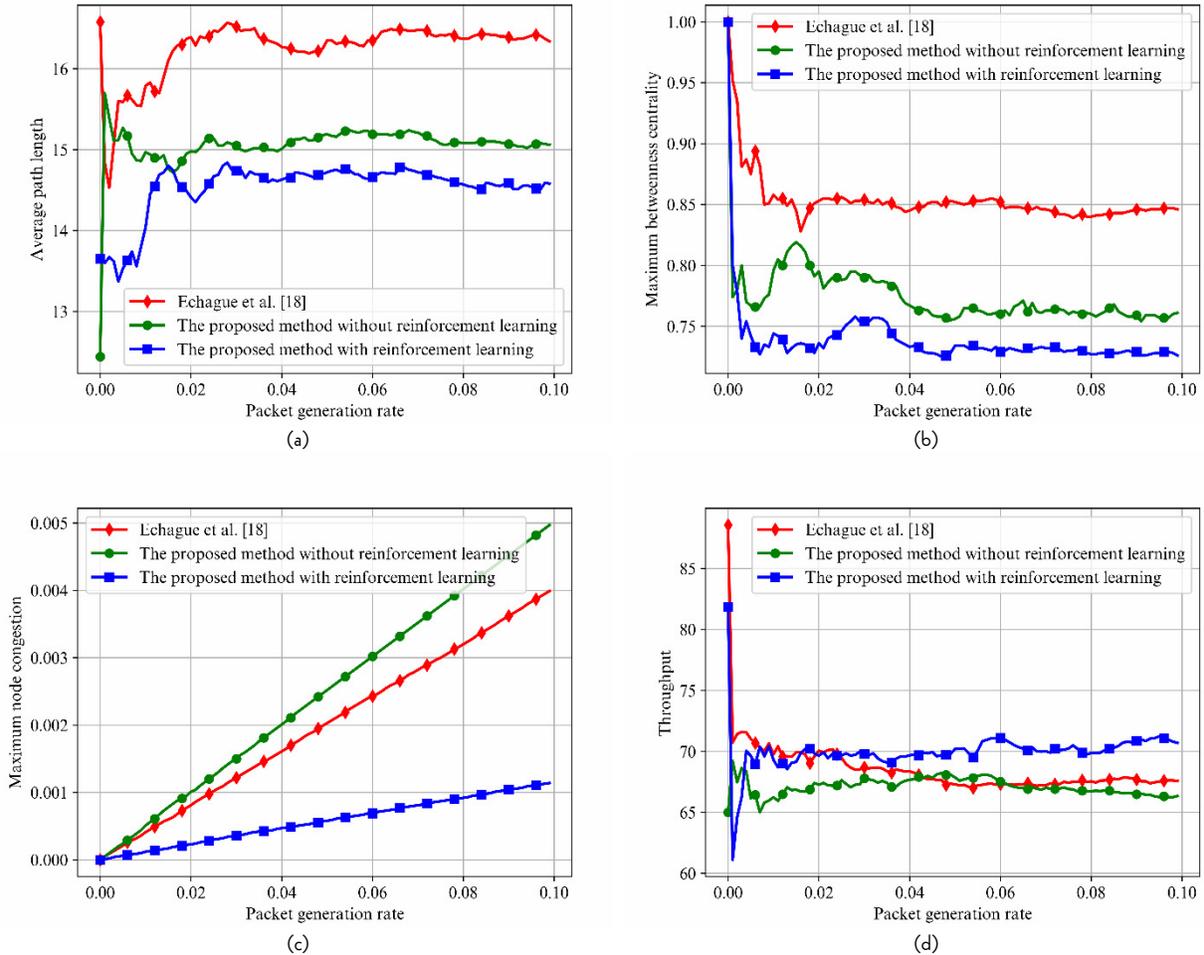

**Figure 8:** (a) Comparison of average path length; (b) maximum betweenness centrality; (c) maximum node congestion, and (d) throughput (percentage) concerning packet generation rate in the network using the Watts-Strogatz small-world model. The assessment includes three different methods under the Poisson traffic pattern, with 64 nodes, an initial regular graph with a degree of 6, and a rewiring probability of 0.5.

## 6. Concluding Remarks

In recent years, extensive research has been dedicated to understanding and mitigating the congestion phenomenon in complex networks. This research primarily falls into three broad categories: soft strategies, hard strategies, and resource allocation strategies. The transmission of packets in a network is intricately linked to its structure and core elements. Routing strategies, the pathways packets take from source to destination, and the management of available resources, such as buffers and communication channels, are crucial aspects when exploring and addressing congestion phenomena. This manuscript proposes a novel routing algorithm employing reinforcement learning to control and manage congestion in complex networks. The proposed scheme has been applied to three network models—Barabási-Albert scale-free networks, Watts-Strogatz small-world model, and Erdös-Rényi random network model—with subsequent evaluations of its effectiveness. Numerical results from simulation experiments indicate that the proposed method can achieve higher efficiency compared to other schemes, even without reinforcement learning. These results suggest potential benefits such as reducing the cost of routing packets by shortening communication paths and enhancing network throughput and capacity.

In this study, we initially strive to identify a suitable approach for managing congestion, one that ensures communication between nodes and utilizes the shortest paths in the network, facilitating packets' shortest routes to their destinations. While this concept aligns closely with the shortest path algorithm, applying such an algorithm directly in the network may induce congestion due to the network's structure. The proposed method, however, effectively distributes the network load evenly among balanced network nodes, preventing congestion in nodes with high betweenness and degree centralities, which are particularly prone to congestion.

A notable challenge for this method is the continuous updating of routes, necessitating real-time assessment of network routes to make optimal decisions based on incoming traffic to network nodes. Updating all routes at each step incurs a significant computational overhead. Consequently, reinforcement learning is employed to expedite updates on routes experiencing premature traffic and critical conditions. Simulations demonstrate the efficacy of this approach in achieving the primary goals of congestion control. The evaluation of average path length across three network models—Barabási Albert, Watts-Strogatz, and Erdös-Rényi—under random and Poisson traffic conditions shows that reducing the average path length delays congestion, balances network load, and diminishes maximum betweenness centrality. Furthermore, judicious route updates enhance network throughput and capacity compared to previous approaches.

Future work in this area could explore optimal resource allocation strategies, considering aspects such as buffer and link management, an area that remains relatively unexplored, especially in the context of machine learning applications. Additionally, given evidence that network traffic may exhibit non-random or Poisson behaviors, particularly in large time scales, there is an opportunity to develop models and methods that evaluate congestion management and control efficiency under self-similar traffic patterns. Studies have shown a significant degradation in network efficiency in the presence of self-similar traffic, prompting the need to provide node structures in complex networks that can support such traffic more effectively.

Moreover, the assumptions in the presented methods presume uniformly distributed destinations for network packets throughout the network, indicating equal probability for packets to be destined for all other nodes. However, certain parallel applications exhibit non-uniform traffic patterns, such as the hot-spot traffic model, where specific nodes receive more packets than others. An extension of this study could involve developing models and methods for managing congestion in complex networks under self-similar traffic with non-uniform destination distribution. Large-scale systems and complex networks often require coordination and interaction among many components, achievable through multicast communication (one-to-many) and broadcasting (one-to-all) within the network. Exploring efficient algorithms for managing and controlling congestion under multicast and broadcast collective communications in complex networks presents another avenue for future research, with potential applications in various real-world scenarios in science and engineering.

# References

[1] W. Huang and T. W. S. Chow, "Investigation of both local and global topological ingredients on transport efficiency in scale-free networks," Chaos Interdiscip. J. Nonlinear Sci., vol. 19, no. 4, p. 043124, Dec. 2009, doi: 10.1063/1.3272217.

[2] A. Arenas, L. Danon, A. Díaz-Guilera, and R. Guimerà, "Local Search with Congestion in Complex Communication Networks," in Computational Science - ICCS 2004, Berlin, Heidelberg, 2004, pp. 1078–1085. doi: 10.1007/978-3-540-24688-6_139.

[3] T. Christiano Silva and L. Zhao, Machine Learning in Complex Networks. Cham: Springer International Publishing, 2016. doi: 10.1007/978-3-319-17290-3.

[4] R. S. Sutton, A. G. Barto, and C.-D. A. L. L. A. G. Barto, Reinforcement Learning: An Introduction. MIT Press, 1998.

[5] L. Buzna and R. Carvalho, "Controlling congestion on complex networks: fairness, efficiency and network structure," Sci. Rep., vol. 7, no. 1, p. 9152, Dec. 2017, doi: 10.1038/s41598-017-09524-3.

[6] T. Panayiotou and G. Ellinas, "Fair Resource Allocation in Optical Networks under Tidal Traffic," in GLOBECOM 2020 - 2020 IEEE Global Communications Conference, Dec. 2020, pp. 1–6. doi: 10.1109/GLOBECOM42002.2020.9322381.

[7] Z.-X. Wu, G. Peng, W.-M. Wong, and K.-H. Yeung, "Improved routing strategies for data traffic in scale-free networks," J. Stat. Mech. Theory Exp., vol. 2008, no. 11, p. P11002, Nov. 2008, doi: 10.1088/1742-5468/2008/11/P11002.

[8] B. Danila, Y. Yu, J. A. Marsh, and K. E. Bassler, "Transport optimization on complex networks," Chaos Interdiscip. J. Nonlinear Sci., vol. 17, no. 2, p. 026102, Jun. 2007, doi: 10.1063/1.2731718.

[9] G. Yan, T. Zhou, B. Hu, Z.-Q. Fu, and B.-H. Wang, "Efficient routing on complex networks," Phys. Rev. E, vol. 73, no. 4, p. 046108, 2006.

[10] X. Zhang, Z. Zhou, and D. Cheng, "Efficient path routing strategy for flows with multiple priorities on scale-free networks," PLOS ONE, vol. 12, no. 2, p. e0172035, Feb. 2017, doi: 10.1371/journal.pone.0172035.

[11] P. Echenique, J. Gómez-Gardenes, and Y. Moreno, "Dynamics of jamming transitions in complex networks," EPL Europhys. Lett., vol. 71, no. 2, p. 325, 2005.

[12] P. Echenique, J. Gómez-Gardeñes, and Y. Moreno, "Improved routing strategies for Internet traffic delivery," Phys. Rev. E, vol. 70, no. 5, p. 056105, 2004.

[13] L. Zhao, Y.-C. Lai, K. Park, and N. Ye, "Onset of traffic congestion in complex networks," Phys. Rev. E, vol. 71, no. 2, p. 026125, 2005.

[14] W.-X. Wang, B.-H. Wang, C.-Y. Yin, Y.-B. Xie, and T. Zhou, "Traffic dynamics based on local routing protocol on a scale-free network," Phys. Rev. E, vol. 73, no. 2, p. 026111, 2006.

[15] P. Holme, "Congestion and centrality in traffic flow on complex networks," Adv. Complex Syst., vol. 6, no. 02, pp. 163–176, 2003.

[16] D. De Martino, L. Dall'Asta, G. Bianconi, and M. Marsili, "Congestion phenomena on complex networks," Phys. Rev. E, vol. 79, no. 1, p. 015101, Jan. 2009, doi: 10.1103/PhysRevE.79.015101.

[17] Y. Yang, H. Zhao, J. Ma, Z. Qi, and Y. Zhao, "An optimal routing strategy on scale-free networks," Int. J. Mod. Phys. C, vol. 28, no. 07, p. 1750087, Jul. 2017, doi: 10.1142/S0129183117500875.

[18] J. Echagüe, V. Cholvi, and D. R. Kowalski, "Effective use of congestion in complex networks," Phys. Stat. Mech. Its Appl., vol. 494, pp. 574–580, 2018.

[19] J. Echagüe, V. Cholvi, and A. Fernández, "Factors affecting congestion-aware routing in complex networks," Phys. Stat. Mech. Its Appl., vol. 587, p. 126483, Feb. 2022, doi: 10.1016/j.physa.2021.126483.

[20] D. J. Watts and S. H. Strogatz, "Collective dynamics of 'small-world' networks," Nature, vol. 393, no. 6684, Art. no. 6684, Jun. 1998, doi: 10.1038/30918.

[21] R. Cohen and S. Havlin, Complex Networks: Structure, Robustness and Function. 2010. doi: 10.1017/CBO9780511780356.

[22] G. Last and M. Penrose, Lectures on the Poisson Process. Cambridge University Press, 2017.